\documentclass[twocolumn]{aastex63}
\usepackage{latexsym}
\usepackage{amssymb}
\usepackage{amsfonts}
\usepackage{amsmath}
\usepackage{bm}
\usepackage{tabularx}
\usepackage[normalem]{ulem}
\usepackage{multirow}
\usepackage{natbib} 
\usepackage{gensymb}
\usepackage{subfigure}
\usepackage{wrapfig}
\usepackage{color}
\usepackage{calc}
\usepackage{amsmath,amssymb,graphicx}
\usepackage{amssymb,amsmath}
\usepackage{tensor}
\usepackage{bm}
\usepackage{pifont}
\usepackage{microtype}
\usepackage{booktabs}
\usepackage{times}
\usepackage[varg]{txfonts}
\usepackage{subfigure}
\usepackage{lipsum}
\usepackage{breakurl} 
\usepackage{graphicx}
\usepackage{pbox}

\definecolor{LinkColor}{rgb}{0.75, 0, 0}
\definecolor{CiteColor}{rgb}{0, 0.5, 0.5}
\definecolor{UrlColor}{rgb}{0, 0, 0.75}
\hypersetup{linkcolor=LinkColor}
\hypersetup{citecolor=CiteColor}
\hypersetup{urlcolor=UrlColor}
\allowdisplaybreaks
\DeclareFontFamily{OT1}{pzc}{}
\DeclareFontShape{OT1}{pzc}{m}{it}{<-> s * [1.10] pzcmi7t}{}
\DeclareMathAlphabet{\mathpzc}{OT1}{pzc}{m}{it}

\newcommand{\be}{\begin{equation}}
\newcommand{\psr}{PSR~J1023$+$0038}
\newcommand{\fgl}{3FGL~J1544.6$-$1125}
\newcommand{\ee}{\end{equation}}
\newcommand{\ber}{\begin{eqnarray}}
\newcommand{\eer}{\end{eqnarray}}
\newcommand{\xss}{XSS~J12270$-$4859~}
\newcommand{\igr}{IGR~J18245$-$2452~}
\newcommand{\rxs}{1RXS~J154439.4$-$112820~}
\newcommand{\cxou}{CXOU~J110926.4$-$650224~}
\newcommand{\fglnew}{4FGL~J0427.8$-$6704~}

\def\bea{\begin{eqnarray}}
\def\eea{\end{eqnarray}}

\accepted{16th August 2021}
\submitjournal{ApJ}

\shorttitle{Radio Continuum Observations of \fgl}
\shortauthors{Jaodand et al.}
\begin{document}

\title{Quasi-Simultaneous Radio/X-ray Observations of the Candidate Transitional Millisecond Pulsar \fgl\ During its Low-Luminosity Accretion-Disc State}

\correspondingauthor{Amruta Jaodand}
\email{ajaodand@caltech.edu}

\author[0000-0002-3850-6651]{Amruta D. Jaodand}
\affiliation{California Institute of Technology, 1200 E Califronia Blvd., Pasadena, CA 91125, USA}
\affiliation{ASTRON, the Netherlands Institute for Radio Astronomy, Postbus 2, 7990 AA Dwingeloo, the Netherlands}
\affiliation{Astronomical Institute Anton Pannekoek, University of Amsterdam, 1098XH, Amsterdam, the Netherlands}

\author[0000-0002-3430-1501]{Adam T. Deller}
\affiliation{Centre for Astrophysics and Supercomputing, Swinburne University of Technology,\\ Mail Number H74, PO Box 218, Hawthorn, VIC 3122, Australia}

\author[0000-0001-6128-3735]{Nina Gusinskaia}
\affiliation{Astronomical Institute Anton Pannekoek, University of Amsterdam, 1098XH, Amsterdam, the Netherlands}
\affiliation{David D Dunlap Institute of Astronomy and Astrophysics,\\ University of Toronto, Ontario, Canada M5S 3H4}

\author[0000-0003-2317-1446]{Jason W. T. Hessels}
\affiliation{ASTRON, the Netherlands Institute for Radio Astronomy, Postbus 2, 7990 AA Dwingeloo, the Netherlands}
\affiliation{Astronomical Institute Anton Pannekoek, University of Amsterdam, 1098XH, Amsterdam, the Netherlands}

\author[0000-0003-3124-2814]{James C. A. Miller-Jones}
\affiliation{International Centre for Radio Astronomy Research - Curtin University, GPO Box U1987, Perth, WA 6845, Australia}

\author[0000-0003-0638-3340]{Anne M. Archibald}
\affiliation{ASTRON, the Netherlands Institute for Radio Astronomy, Postbus 2, 7990 AA Dwingeloo, the Netherlands}
\affiliation{Astronomical Institute Anton Pannekoek, University of Amsterdam, 1098XH, Amsterdam, the Netherlands}
\affiliation{School of Mathematics, Statistics, and Physics, Newcastle University, NE17RU UK}

\author[0000-0002-9870-2742]{Slavko Bogdanov}
\affiliation{Columbia Astrophysics Laboratory, Columbia University, 550 West 120th Street, New York, NY, 10027, USA}

\author[0000-0002-1429-9010]{Cees Bassa}
\affiliation{ASTRON, the Netherlands Institute for Radio Astronomy, Postbus 2, 7990 AA Dwingeloo, the Netherlands}

\author[0000-0002-3516-2152]{Rudy Wijnands}
\affiliation{Astronomical Institute Anton Pannekoek, University of Amsterdam, 1098XH, Amsterdam, the Netherlands}

\author[0000-0002-6459-0674]{Alessandro Patruno}
\affiliation{Institute of Space Sciences (IEEC-CSIC) Campus UAB, Carrer de Can Magrans, s/n, 08193 Barcelona, Spain}
\affiliation{ASTRON, The Netherlands Institute for Radio Astronomy, Postbus 2, 7900 AA Dwingeloo, The Netherlands}

\author[0000-0002-5956-5546]{Sotiris Sanidas}
\affiliation{Jodrell Bank Centre for Astrophysics, School of Physics and Astronomy, The University of Manchester, Manchester M13 9PL,UK}


\begin{abstract}

%Currently 250 words - limit is 250
\fgl\ is a candidate transitional millisecond pulsar (tMSP).  Similar to the well-established tMSPs --- \psr, \igr\ and \xss\ --- \fgl\ shows $\gamma$-ray emission and discrete X-ray `low' and `high' modes during its low-luminosity accretion-disc state.  %\psr\ is much brighter in radio than previously expected for neutron star systems accreting at low X-ray luminosities ($L_{\rm 2-10 keV} \sim 10^{33}$\,erg\,s$^{-1}$), and is comparably radio-bright to stellar-mass black hole binary systems at similar X-ray luminosity.  
Coordinated radio/X-ray observations of \psr\ in its current low-luminosity accretion-disc state showed rapidly variable radio continuum emission --- possibly originating from a compact, self-absorbed jet, the propellering of accretion material, and/or pulsar moding.  \fgl\ is currently the only other (candidate) tMSP system in this state, and can be studied to see if tMSPs are typically radio loud compared to other neutron star binaries. In this work, we present a quasi-simultaneous VLA and \textit{Swift} radio/X-ray campaign on \fgl.  We detect 10-GHz radio emission varying in flux density from $47.7 \pm 6.0$\,$\mu$Jy down to $\lesssim 15$\,$\mu$Jy ($3\sigma$ upper limit) at four epochs spanning 3 weeks.  At the brightest epoch, the radio luminosity is $L_{\rm 5GHz} = 2.17\pm0.17~\times10^{27}$\,erg\,s$^{-1}$ for a quasi-simultaneous X-ray luminosity $L_{\rm 2-10 keV} = 4.32\pm0.23~\times10^{33}$\,erg\,s$^{-1}$ (for an assumed distance of 3.8\,kpc).  These luminosities are close to those of \psr, and the results strengthen the case that \fgl\ is a tMSP showing similar phenomenology to \psr.  %Strictly simultaneous radio/X-ray observations are needed to determine whether, like \psr, the radio emission from \fgl\ brightens during X-ray low modes. 
\end{abstract}



\section{Introduction} \label{sec:intro}

The transitional millisecond pulsars (tMSPs) are a recently discovered type of binary system that transition between rotation-powered radio millisecond pulsar (RMSP) and accretion-powered low-mass X-ray binary (LMXB) states \citep[e.g.,][]{ASR:2009,PFB:2013,SAH:2014,BPH:2014,RRB:2015}. Currently, there are three well-established tMSP systems: \psr, \igr and \xss and three excellent candidate transitional millisecond pulsar systems \fgl,
\cxou \citep{ZPM:2019} and \fglnew \citep{SLC:2016,KBC:2020}.

To conclusively classify a neutron star binary as a tMSP, one needs to demonstrate that the system has been observed in both the rotation-powered RMSP and accretion-powered LMXB states.  Nonetheless, in the absence of an observed transition, we can still make a strong case that a particular candidate system is of a similar ilk to the tMSPs and may in the future undergo a state transition.  Fortunately, the tMSPs provide a rich observational phenomenology that helps build such a connection (see, Table \ref{table:obssum}), with the caveat that there are only three such systems currently known and perhaps a larger sample will display a larger diversity. %moved here from the conclusion on account of referee request

In their LMXB\footnote{In this paper we use the term `LMXB state' to distinguish from when a tMSP is observed as an RMSP, but caution that only one confirmed tMSP has been observed in a full LMXB outburst, whereas the other two have only been seen in a low-luminosity, persistent accretion-disc state that is unlike what has been observed to date in classical LMXBs.} state, tMSPs show distinct X-ray luminosity modes.  For example, \psr\ switches within $\lesssim 10$\,s between `high' ($L_{\rm 2-10 keV} \sim 10^{33}$\,erg\,s$^{-1}$, present $\sim$80\% of the time) and `low' ($L_{\rm 2-10 keV} \sim 10^{32}$\,erg\,s$^{-1}$, present $\sim$20\% of the time) X-ray luminosity modes \citep{ABP:2015, BAB:2015, JAH:2016}. These modes last typically for minutes up to an hour.  Occasionally, there are instances of X-ray `flares' ($L_{\rm 2-10 keV} \sim 10^{34}$\,erg\,s$^{-1}$). A very similar variability pattern was observed in \xss, prior to its 2012 transformation to an RMSP \citep{deMartino:2013}, whereas \igr's moding behaviour differs as it shows longer-lasting modes \citep[several hour duration; see][]{Linares:2014}. 

Coherent X-ray pulsations are observed  in \psr\ and \xss\ at the neutron star's rotational period during only the X-ray high mode \citep{ABP:2015,BAB:2015,PMB:2015,JAH:2016, PAP:2019}. Coherent pulsations at X-ray luminosities typically considered as `quiescent state' for other X-ray binaries is surprising since the accretion material is unlikely to provide enough pressure to overcome the neutron star's magnetic pressure and to produce hot spots on the stellar surface --- as is believed to powered coherent pulsations from accreting X-ray millisecond pulsars (AMXPs). The effect has thus far only been observed in tMSPs, and could represent some exotic form of stimulated pulsar moding \citep{ABP:2015}.   
This moding behaviour now extends to to optical wavelengths \citep[see,]{PAP:2019}. 

\psr\ has been monitored across the electromagnetic spectrum since its transition to  the LMXB state in 2013 \citep{PAH:2014,SAH:2014}.  In the LMXB state, \psr\ is the best studied tMSP system. Hence, we use it hereafter as the canonical system to outline phenomenological similarities in the tMSP class.

To better understand the puzzling low-luminosity tMSP accretion regime, we conducted a timing campaign on \psr\ using \textit{XMM-Newton} \citep{JAH:2016}.  In this two-yr-long campaign we timed the coherent X-ray pulsations
%\footnote{This analysis was done after fitting for stochastic orbital variations by varying the time of ascending node at each epoch.} 
and found that the system is spinning down $26.8\pm0.4\%$ faster than in the RMSP state \citep[i.e. compared to the radio-derived timing solution of][]{AKH:2013}. This measurement shows that the accretion torques have (on average) been roughly constant over 2\,yr.  Moreover, the lack of large torque variations, coupled with the spin down rate being only $\sim 27$\% faster than in the RMSP state, tells us that the dipole spin-down mechanism observed in the RMSP state is still dominant and the pulsar wind continues to operate in a mostly unchanged manner. Enhanced spin-down could in principle be due to an increased proportion of open field lines in the pulsar magnetosphere \citep{PSB:2016,PSB:2017} or from outflow of accretion material.

Additional insights come from radio continuum imaging of \psr\ using the Karl G. Jansky Very Large Array (VLA).  \citet{DMM:2015} found approximately flat-spectrum ($-0.3 < \alpha < 0.3$, where the radio flux density $S$ scales with the observing frequency $\nu$ as $S \propto \nu^{\alpha}$) continuum radio emission that was observed to vary by up to an order-of-magnitude in brightness on timescales of minutes to an hour. This suggested partially self-absorbed synchrotron emission from a jet-like collimated, outflow. The radio emission, which persisted for the span of the 6-month observing campaign, was surprisingly bright given the low X-ray luminosity and inferred accretion rate.  This was attributed to radiatively inefficient `propeller-mode' accretion, where there is ejection of inflowing accretion material as it is unable to overcome the centrifugal barrier at the co-rotation radius (where accretion material co-rotates with the neutron star's magnetic field). In such a situation, the jet-like outflow would then carry away the majority of the liberated accretion energy.

%Also, the current neutron star radio-jet-formation models have been shown to differ from observational evidences such as unexpected jets found in high magnetic field NS-LMXB \citet{EDR:2018} and 

The tMSPs hence accrete at unexpectedly low X-ray luminosity levels \citep{ABP:2015} and show higher radio luminosities than expected for other NS-LMXBs with millisecond spin rates \citep{DMM:2015}. For example, Z-sources \footnote{They trace a Z-shape comprising of flaring, normal and horizontal branches from harder to softer colors in the hardness-intensity diagram (HID).} are highly luminous as they accrete near-Eddington luminosities ($\geq~10^{38}$\,ergs s$^{-1}$), and their outbursts span timescales of a $\sim$~few hrs to days. Atoll-sources\footnote{Their HID traces show steady hardness over range of intensities resembling small clusters or ``islands"  \citep[see,][]{HK:1989,Muno_2002,CGB:2014}} are usually found to be accreting at luminosities $\leq~10^{38}$\,ergs s$^{-1}$ on month-long timescales. Finally, rapid burster are peculiar neutron star binary which shows Type-I X-ray burst at the highest peak of outburst and the rare,intense Type-II X-ray bursts \footnote{Recent works by \citet{DMH:2014,EBD:2017} indicate that dramatic type-II X-ray bursts occur due to a gap between neutron star and inner edge of the accretion disk.} erratically during periods of lower X-ray activity. The stable nature of the long-lasting (multi-year) accretion regime and abundance of information now available from multi-wavelength campaigns \citep[e.g.,][]{BAB:2015} has established tMSPs as excellent laboratories to study accretion onto magnetized neutron stars, at lower accretion rates than probed in previous studies. 

Most recently, \citet{BDM:2018} uncovered a striking radio/X-ray anti-correlation in \psr\ using strictly simultaneous \textit{Chandra X-ray Observatory} and VLA observations. They found that all the X-ray `low' modes were reliably accompanied by simultaneous radio flux brightening, whose duration was closely matched to that of the low mode.  Additional, sporadic radio flares -- unassociated with X-ray low/high-mode switches, and only sometimes associated with X-ray flares -- were also occasionally seen.  During the X-ray low-mode, the radio spectral index was observed to change from significantly inverted ($\alpha \simeq 0.4$) in the radio flare rise to relatively steep ($\alpha \simeq -0.5$) during the radio flare decay. Going beyond the initial insights from the \citet{DMM:2015} radio/X-ray campaign, \citet{BDM:2018} concluded that the observations could not be explained by a canonical inflow/outflow model, in which the radiated emission and the jet luminosity are powered by, and positively correlated with, the available accretion energy. They argued instead for a picture in which the X-ray high modes (with accompanying coherent X-ray pulsations) represent active accretion into the neutron star magnetosphere \citep[as also concluded by][]{ABP:2015}, whereas the X-ray low modes and associated radio flares represent the rapid ejection of plasma by the active rotation-powered pulsar -- possibly initiated by a reconfiguration of the neutron star magnetosphere to another quasi-stable state.  The only slightly enhanced neutron star spin-down rate found by \citet{JAH:2016} appears consistent with this picture, in which the rotation-powered pulsar wind continues to play a significant role in shaping the multi-wavelength emission properties of the system.

The cumulative observational findings achieved in the last decade of tMSP studies \citep{JHA:2018, CS:2018, pd:2020} have given various insights into low-level accretion onto rapidly-spinning, highly magnetised neutron stars. We recap these observational findings for various sources in Table \ref{table:tabsum}. The comprehensive picture of this accretion state, however, necessitates an increase in the population of known tMSPs, especially since two of the three known systems (\xss\ and \igr) are currently in the RMSP state.

\begin{deluxetable*}{c c c c}[ht]
\centering
\tablenum{1}
\tablecaption{Summary of tMSP observational properties\label{table:tabsum}}
\tablewidth{0pt}
\tablehead{
\colhead{Properties} & \colhead{\psr} & \colhead{\igr} & \colhead{\xss}}
\startdata
\hline
& Radio Millisecond Pulsar State & &\\
\hline
Steep-spectrum radio pulsations& \ding{51} &\ding{51}&\ding{51} \\
Gamma-ray pulsations& \ding{53} &\ding{51}&\ding{51}\\
Orbitally modulated optical and X-ray emission& \ding{51} &\ding{51}&\ding{51} \\
Stochastic orbital variations& \ding{51} &\ding{51}&\ding{51} \\
Radio Eclipses & \ding{51} &\ding{51}&\ding{51}\\
\hline
& Low-mass X-ray Binary State & &\\
\hline
Flat spectrum radio emission & \ding{51} &\ding{51} &\ding{51} \\
Gamma-ray flux enhancement & \ding{51} &\ding{51} &\ding{51}\\
X-ray flux enhancement and pulsations & \ding{51} &\ding{51} &\ding{51} \\
Stochastic orbital variations & \ding{51} & \ding{51} &\ding{51} \\
\enddata
\tablecomments{Summary of key observational properties of tMSPs in RMSP and LMXB states.}
\end{deluxetable*}

\subsection{\fgl}

A very promising candidate tMSP, \fgl, was recently found by \citet{BH:2015}. Motivated by the fact that \psr\ underwent a factor of $5$ brightening at GeV energies when it transitioned from an RMSP to an LMXB state \citep{SAH:2014}, this candidate tMSP was found by investigating the population of unidentified \textit{Fermi} gamma-ray sources. More broadly, the selection criteria were based on the observed gamma-ray and multi-wavelength emission from tMSPs in both RMSP and LMXB states \citep{PAH:2014, TRC:2015}. \citet{BH:2015} used the optical properties and X-ray light curve phenomenology to categorise \fgl\ (also associated with X-ray source \rxs) as a strong tMSP candidate. 

More extensive studies, which analysed the archival X-ray and radio data \citep{Bog:2016} for this source, further strengthened this claim and showed that the system had not transitioned to an RMSP state over the course of the past ten years (at least not for a prolonged period of time). Also, using spectroscopic observations \cite{BSC:2017} showed that the orbital period of \fgl\ is $\sim 0.24$\,day, similar to other tMSPs ($P_{orb} = 0.2-0.4$\,day), and that it is situated at $3.8\pm0.7$\,kpc. 

Besides \psr, \fgl\ at the time of its discovery was the only known (candidate) tMSP system in the LMXB state.  Thus, in order to capitalize on this rare opportunity to study another tMSP in the LMXB state, we undertook a dedicated quasi-simultaneous radio/X-ray observation campaign using the VLA and the \textit{Neil Gehrels Swift Observatory} X-Ray Telescope (XRT). In this paper, we present the results of this campaign. In Section 2 we present the observations and analysis. In Section 3 we present the results and in Section 4 the astrophysical implications derived from them.

\label{sec:obs}

\subsection{Very Large Array Observations}

After the discovery of \fgl\ as a strong tMSP candidate, we observed the source using the upgraded VLA \citep{PCB:2011} on four different occasions (2015 May 12, 15, 30, and June 5) under the Director's Discretionary Time project 15A-484 (PI: Jaodand; legacy ID: AJ401). 
Since our goal was to determine the location of \fgl\ on the radio/X-ray correlation plane, our VLA observations were dynamically scheduled to be at least quasi-simultaneous with \textit{Swift} XRT observations (Figure~\ref{fig:qs})\footnote{The data from these observations is publicly accessible from the National Radio Astronomy Observatory (NRAO) VLA archive. \url{https://archive.nrao.edu/archive/archiveproject.jsp}}.

The scheduling block duration for each observation was one hour, and we used the X-band ($8-12$\,GHz) receiver (using the standard wide-band continuum correlator set-up with 3-bit sampling). The VLA was in an evolving hybrid `BnA' configuration during our observations, and offered a resolution of $\sim$1\arcsec\ at 10\,GHz.  In all but the third observation, some antennas were in the process of moving into or out of the BnA configuration, and so the sensitivity and $uv$-coverage were slightly compromised due to a reduced number of available antennas in the array.

In each epoch, the target source \fgl\ was observed for a total time of $\sim$32 min (10 scans of $\sim$3.2 minutes). The source J1331$+$305 (3C286) was used as a flux calibrator and to calibrate the instrumental bandpass; flux density scale calibration coefficients were adopted from \citet{Perley_2017}. In all the observations, J1543$-$0757 was used as a gain calibrator and was visited every $\sim$4\,minutes, in between target scans. For the third and fourth epochs, a scan on the source J1407$+$2827 was also added to allow for the potential calibration of the instrumental polarization leakage. We summarize these observations in Table \ref{table:obssum}.

\subsection{VLA Data Analysis}

Every VLA scheduling block between 2013 January and 2015 September was automatically flagged and calibrated by a VLA scripted calibration pipeline\footnote{\url{https://science.nrao.edu/facilities/vla/data-processing/pipeline/scripted-pipeline}}. This pipeline uses CASA (v. 5.1.2) as the underlying software suite \citep[for more information see][]{MWS:2007} to handle various operations such as loading the data into a measurement set, Hanning smoothing, flux calibration, radio frequency interference (RFI) flagging, etc. 

Based on an inspection of the pipeline data products for each epoch, we identified additional RFI and antenna/frequency combinations affected by instrumental issues, and performed manual flagging to remove the affected data.  Finally, we re-ran the pipeline to generate the best possible calibration, and applied this to the \fgl\ field before splitting the data and imaging the target field.

The brightest source in the field is located several arcminutes to the northwest of \fgl, and has a flux density of $\sim$300\,$\mu$Jy (after correcting for attenuation by the primary beam).  The limited brightness of this source proved insufficient to allow for the derivation of self-calibration solutions, and so no further calibration was applied to the target field.

For each epoch, we used the CASA task {\tt clean} in interactive mode to image the full primary beam of the VLA in Stokes I (total intensity), identifying and cleaning sources above 30\,$\mu$Jy ($\gtrsim$5$\sigma$ compared to the noise level).  We used the full bandwidth of the observations and a simple clean (natural weighting, multi-frequency synthesis with a single Taylor term and a single scale).

In the resulting images, we estimated the rms noise using a source-free box near the position of \fgl.  If a peak $>3\sigma$ was present at the known position of \fgl\ \citep{BH:2015}, we used the task {\tt imfit} to fit a single 2-D Gaussian component (with semi-major and semi-minor axes matching the size of the synthesized beam) and took the fitted value of the peak as the flux density of the point source at that epoch. For the epochs where \fgl\ was not detected, we report a $3\sigma$ upper limit based on the off-source rms noise.  We include a 10\% uncertainty to account for the absolute flux calibration of the VLA \citep{PB:2017} in addition to the formal uncertainties obtained from the image rms noise, although the latter dominates.

%\fgl\ was found to be faint, with the strongest detection being $\sim8\sigma$ in significance, and two epochs only providing upper limits on the radio flux density (Table~\ref{table:obssum}). Accordingly, we did not attempt to fit for the spectral index of \fgl, nor did we perform polarization imaging, as neither would offer useful constraints.  Likewise, testing for variability is strongly limited by the source brightness and the fact that the low modes -- where, in analogy to \psr, we expect radio brightening -- typically last for only a few minutes to half an hour at a time \citep{BH:2015}.  For reference, the expected rms noise for 5-minute sub-integrations is $\sim 15$\,$\mu$Jy/beam, compared to the $\sim 5$\,$\mu$Jy/beam that we achieve in our $\sim 40$ minutes of on-source time per epoch.  As such, variations of roughly $5-10\times$ compared to the mean brightness would need to occur in order to reliably detect variability on few-minute timescales.  Consequently, we plan to better investigate short-timescale ($< 1$\,hr) variability in a strictly simultaneous radio/X-ray observation campaign, where the X-ray low/high-mode separation can guide the investigation of the radio light curve.

\begin{deluxetable*}{c c c c c c}[ht]
\centering
\tablenum{2}
\tablecaption{Summary of VLA Radio Observations\label{table:obssum}}
\tablewidth{0pt}
\tablehead{
\colhead{Epoch \#} & \colhead{Obs. Date} & \colhead{Obs. Start Time} & \colhead{Obs. Start (MJD)} & \colhead{Average Flux Density} & \colhead{Uncertainty}\\
&\colhead{UTC} & \colhead{UTC} & & \colhead{$\mu$Jy} & \colhead{$\mu$Jy}}
\startdata
1&2015-05-12& 03:29:10 & 57154.17 &23.6 & 4.8\\
2&2015-05-15& 03:10:26 & 57157.15 &47.7 & 6.0\\
3&2015-05-30 & 01:56:23& 57172.10 &$<$13.8 & ---\\
4&2015-06-05 & 08:19:40& 57178.36 &$<$16.2 & ---
\enddata
\tablecomments{{Summary of VLA observations at $8-12$\,GHz (X band) in BnA array configuration.  At each epoch we achieved roughly 40 minutes on source and an imaging rms noise of $\sim 5$\,$\mu$Jy/beam, as expected given the observational setup.  \fgl\ was only convincingly detected in Epochs 1 and 2, where we used a 2-D Gaussian fit to measure the flux density and quote a 1-$\sigma$ uncertainty equivalent to the off-source rms noise.  In Epochs 3 and 4, no significant radio emission is detected at the known source position; here we quote 3-$\sigma$ upper limits based on the off-source rms noise.}}
\vspace{-0.5 in}
\end{deluxetable*}

\subsection{\textit{Swift} Observations}
Prior to our VLA radio campaign, \fgl\ was already being monitored regularly, along with a few `redback' pulsars\footnote{These are a class of eclipsing radio MSP to which all known tMSPs belong when they are in the RMSP state \citep[see, e.g.,][]{JHA:2018}.} and known tMSP systems, using \textit{Swift} XRT \citep{BHN:2004} in $2015$.  The aim of that campaign was to catch any possible transition in \fgl\ from the LMXB to RMSP state -- or at least to establish whether it shows long-term X-ray brightness variations.  We supplemented this campaign with an additional set of \textit{Swift} XRT observations, scheduled close in time to complement the VLA observations of \fgl.

\begin{figure*}
\begin{center}
\includegraphics[width=0.7\textwidth,height=0.4\textwidth]{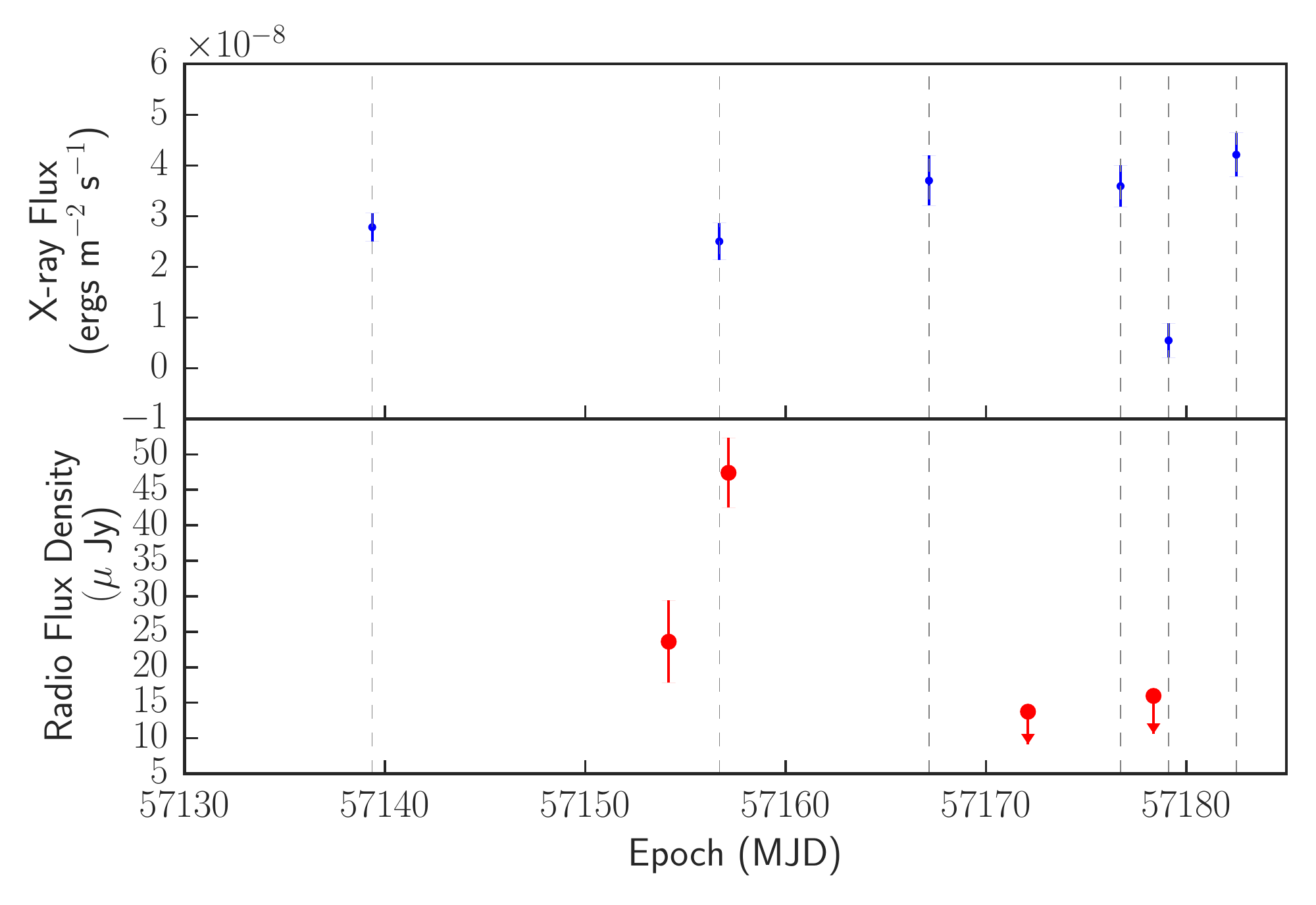}
\caption{Radio flux density at 10\,GHz (red points; bottom panel; see Table~\ref{table:obssum}) and $1-10$\,keV unabsorbed X-ray flux (blue points; top panel; see Table~\ref{table:Swift_obs}) as a function of modified Julian day (MJD). The \textit{Swift} X-ray observations span the four-epoch VLA radio campaign, and the exact epochs of \textit{Swift} observations are denoted by dashed vertical lines for reference. Note that the fifth X-ray observation shown here (at MJD~57179) was only a 418-s exposure, and it is possible that \fgl\ was in the X-ray low-mode for the entirety of this observation, which would explain the low measured X-ray flux compared to the other \textit{Swift} observations. The VLA flux density upper limits ($3\sigma$) at Epochs 3 and 4 are marked with downward arrows.
\label{fig:qs}} 
\end{center}
\end{figure*}

\begin{table*}
\tablenum{3}
\hspace{-0.4 in}
\caption{\normalsize{Summary of \textit{Swift} XRT X-ray Observations }}
\begin{center}
\begin{tabular}{c c c c c}
\hline
&&&\\[-1 em]
Obs. ID & Obs. Date & Flux ($1-10$\,keV) & Uncertainty & Exposure Time\\ 
        & (MJD)     & {\scriptsize ($\times 10^{-12}$\,ergs\,cm$^{-2}$\,s$^{-1}$)} & {\scriptsize ($\times 10^{-12}$\,ergs\,cm$^{-2}$\,s$^{-1}$)} & (ks)\\
&&&\\[-1 em]
\cline{1-5}
\hline
&&&\\[-1 em]
00036065001    &    53983.53686    &    1.74    & 0.61 & 0.6\\
00036065002    &    54117.51016    &    3.52    & 0.17 & 6.2\\
00036065003    &    55966.43553    &    4.58    & 0.23 & 5.1\\
00036065004    &    56091.78215    &    3.32    & 0.31 & 1.9\\
\hline
00036065005    &    57113.40175    &    3.28    & 0.31 & 2.1\\
00036065006    &    57139.36776    &    2.78    & 0.28 & 1.7\\
00036065007    &    57156.69384    &    2.50    & 0.36 & 1.9\\
00036065008    &    57167.16071    &    3.70    & 0.49 & 0.9\\
00036065009    &    57176.72122    &    3.59    & 0.41 & 0.8\\
00036065010    &    57179.12038    &    0.55    & 0.34 & 1.1\\
00036065011    &    57182.49995    &    4.21    & 0.43 & 0.4\\
00036065012    &    57185.89694    &    2.54    & 0.36 & 1.2\\
00036065015    &    57199.32349    &    1.94    & 0.60 & 1.0\\
00036065016    &    57203.88762    &    3.48    & 0.29 & 0.2\\
00036065017    &    57204.35402    &    3.79    & 0.33 & 0.4\\
00036065018    &    57223.57096    &    2.63    & 0.44 & 2.1\\
00036065020    &    57254.26046    &    3.88    & 0.66 & 1.8\\
\hline
\end{tabular}
\\[1.5ex]
\tablecomments{The quoted fluxes are corrected for absorption; see main text.  The uncertainties are at the 90\% confidence level.}%\textcolor{blue}{CHECK!- Ask Slavko}}
\label{table:Swift_obs}
\end{center}
\end{table*}

\begin{figure*}
\begin{center}
\includegraphics[width=1\textwidth]{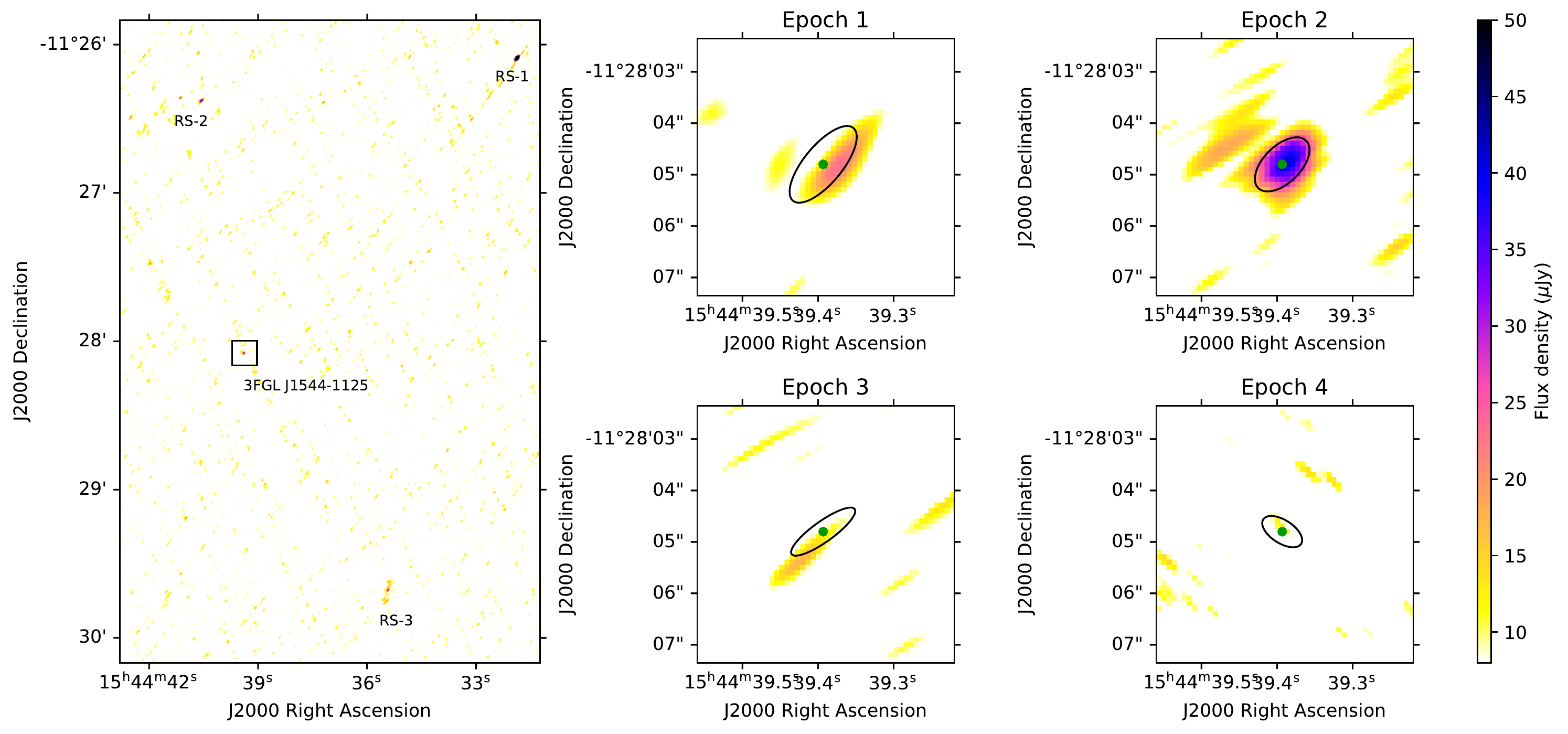}
\caption{Radio interferometric images (Stokes I) from all four VLA observation epochs (see Table~\ref{table:obssum}). {\bf Left:} Radio image of the second epoch showing \fgl\ and three unidentified radio sources in the surrounding field (RS-1, RS-2, RS-3 with the corresponding flux densities of $\sim$350 $\mu$Jy/beam, $\sim$50 $\mu$Jy/beam, $\sim$40 $\mu$Jy/beam). {\bf Right:} Each sub-panel shows zoomed-in radio images of \fgl\ from all 4 epochs. The synthesized beams are shown by black ellipses centered at the Gaia position of the source, which is shown by a green circle. The source is not convincingly detected during Epoch 3 or 4.}
\label{fig:radiobright} 
\end{center}
\end{figure*}

\begin{figure*}
\begin{center}
\includegraphics[width=0.9\textwidth]{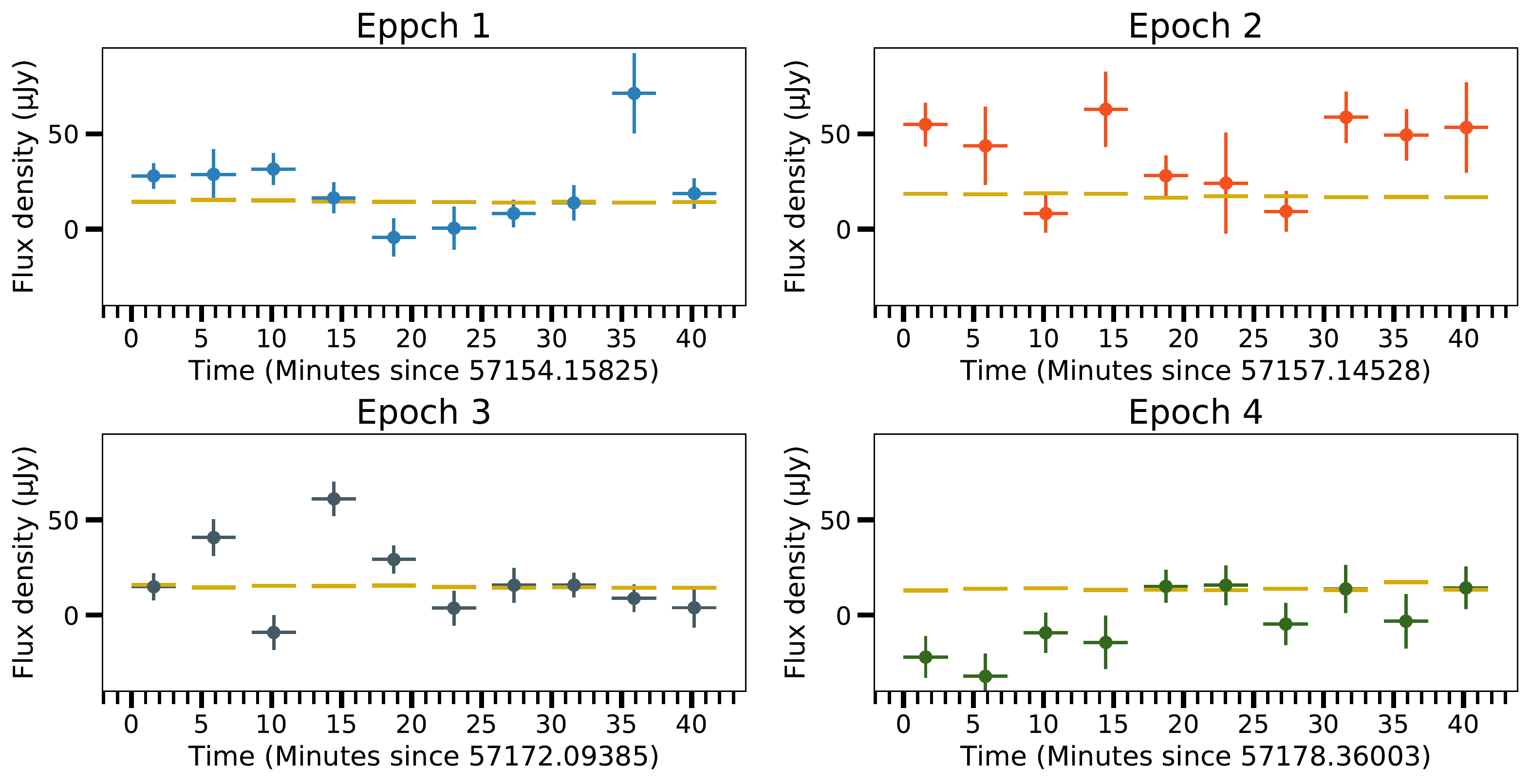}
\caption{Short time-scale radio light-curves constructed over sub-integrations of 3.2 mins, for each VLA observation epoch (see Table~\ref{table:obssum}). Different panels represent corresponding epochs.
The yellow dashed line shows rms flux density levels for each sub-integration. Epoch 2 is overall the brightest although excursions to similar flux levels are seen for short intervals during Epoch 1 and Epoch 3.}
\label{fig:radiolcs} 
\end{center}
\end{figure*}

These observations, as well as four archival observations from $2006-2012$, are summarized in Table~\ref{table:Swift_obs}.  We present $13$ \textit{Swift} observations from 2015 (each of which has a typical duration of $\sim 0.5-2$\,ks), of which six closely span the duration of the VLA observations. 
Although not strictly simultaneous with the VLA observations, these cover the baseline of VLA observations in a quasi-simultaneous manner and demonstrate that the system was still in the LMXB state at that time (Figure~\ref{fig:qs}).
We note that \textit{Swift} has continued to monitor this source and that other observations -- in addition to those presented here, and including the UVOT instrument -- also exist.

\subsection{\textit{Swift} Data Analysis}

We analyzed all \textit{Swift} XRT observations, taken in the photon counting (PC) mode, using the standard \textit{Swift} tool \texttt{xrtpipeline} in HEASoft\footnote{Available at \url{https://heasarc.nasa.gov/lheasoft/}}. Photon counts were extracted from a circular region with a radius of 20 detector pixels (corresponding to 47$''$). We calculated the contribution from the background by averaging counts detected in 4 circular regions of the same size across the field of view after verifying that they did not fall close to obvious point sources. The 1$-$10\,keV X-ray counts were then extracted and binned in 10-s intervals with the software \texttt{xselect}. The X-ray counts were then corrected with the task \texttt{xrtlccorr}, which accounts for telescope vignetting, point-spread function corrections and bad pixels/columns and re-binned the counts in intervals of 50 seconds.
%}
 
%**\textcolor{blue}{placeholder for more analysis from Slavko}

A summary of all the observations and associated flux measurements is presented in Table~\ref{table:Swift_obs}.  The unabsorbed X-ray fluxes (1-10\,keV) for \textit{Swift} XRT observations were obtained assuming a simple absorbed power-law spectral model, as measured using \textit{XMM-Newton} and described in \citet{BH:2015}. We adopted their values for photon index $\Gamma=1.68 \pm 0.04$ and a hydrogen column density of $N_{H}=(1.4 \pm 0.1) \times 10^{21}$\,cm$^{-2}$ along the line of sight.

\begin{figure*}[!ht]
\begin{center}
\includegraphics[width=0.95\textwidth]{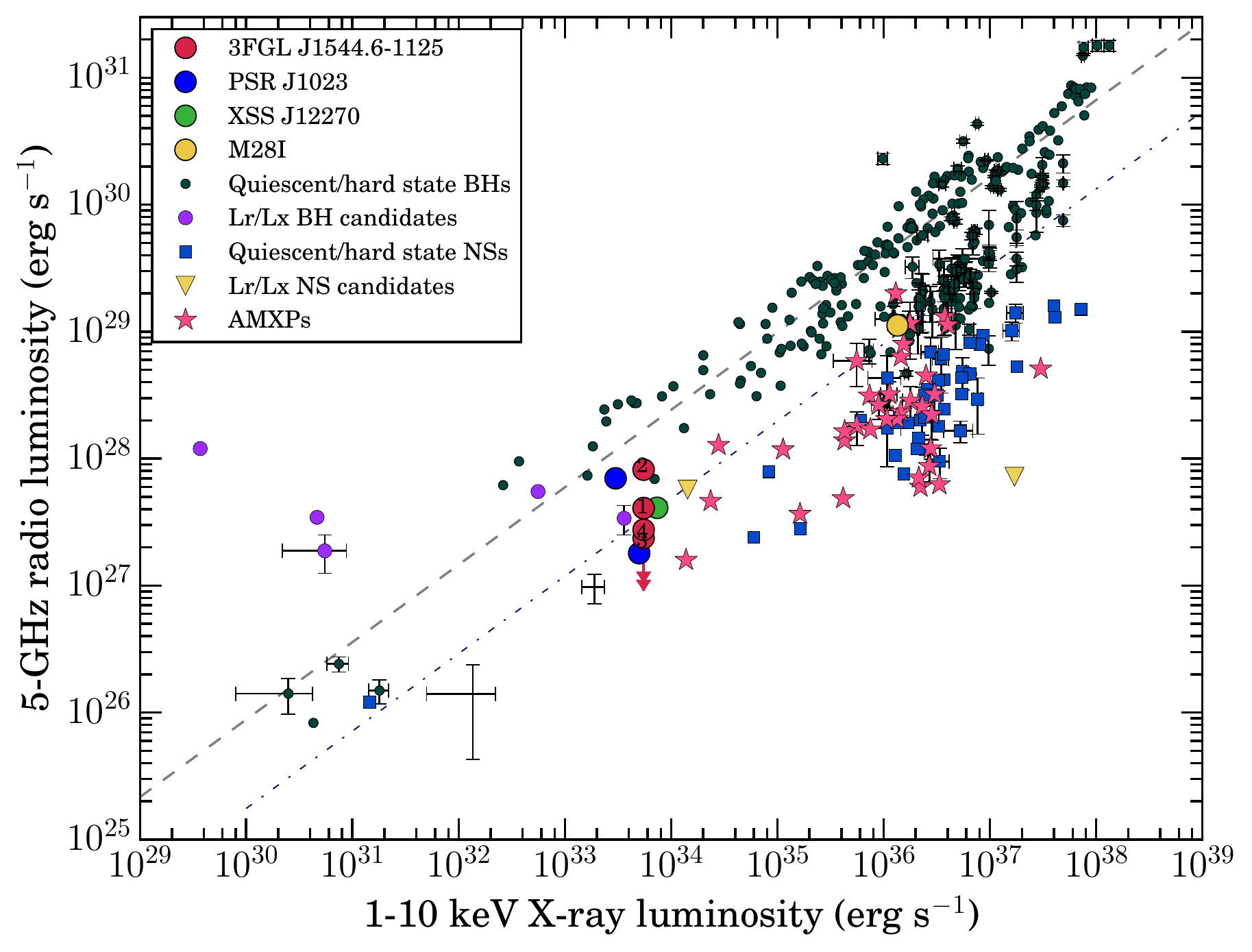}
\caption{A phase space for NS- and BH-LMXBs, showing their 5-GHz radio luminosities versus their $1-10$\,keV X-ray luminosities, as derived from quasi-simultaneous observations in both bands.  Dark violet circles, blue squares and pink stars represent stellar-mass black hole binaries, hard-state neutron star X-ray binaries and accreting millisecond X-ray pulsars, respectively \citep[the values are obtained from the database maintained by][]{Arash:2018}. The thin grey dashed lins shows the best-fit radio/X-ray luminosity correlation for BH-LMXBs \citep{GMR:2014}. The dot-dashed blue line is the putative $L_R \propto {L_X}^{0.7}$ scaling for tMSPs \citep{DMM:2015}. Note that individual systems, with evolving accretion rates, may be represented by multiple points on this diagram.  Previous radio/X-ray observations of individual tMSPs are also shown: M28I/\igr\  \citep[yellow circle;][]{PFB:2013}, \xss\ \citep[green circle;][]{HSC:2011}, and \psr\ \citep[blue circles;][]{BDM:2018}.  For \psr, the radio/X-ray luminosities are shown separately for the low (radio brighter) and high (radio fainter) X-ray modes -- as derived from strictly simultaneous radio and X-ray observations.  The four radio observation epochs of \fgl\ are shown individually using red circles (see also Table~\ref{table:obssum}).  For all four epochs we use a median value for X-ray luminosity of 5.51~$\times10^{33}$\,erg\,s$^{-1}$.  The $3\sigma$ upper limits at Epochs 3 and 4 are indicated with an additional downward arrow.}
\label{figure:lrlx}
\end{center}
\end{figure*}

\section{Results}
Throughout the paper, we assume the spectroscopically estimated distance of $3.8 \pm 0.7$\,kpc for \fgl\ \citep{BSC:2017}.

The \textit{Swift} XRT flux measurements from MJD~53983$-$57254 (2006 Sep to 2015 Aug) are consistent with \fgl\ being in the LMXB state for this entire time. However, given that the state transition to RMSP state can be rapid \citep[e.g.][]{SAH:2014} and might only last for months to years in some cases, we cannot rule out that \fgl\ underwent a switched states during the sparse X-ray monitoring available before 2015.  Regardless, it is clear that the source was in the LMXB state during our 2015 May to Jun VLA campaign: we find an average X-ray luminosity of $4.32\pm0.23~\times10^{33}$\,erg\,s$^{-1}$ at this time, which we use as the point of reference for our radio measurements. While the X-ray flux here was lower on MJD~57179, it is plausible that this observation was dominated by a single low-mode in this short (418-s) observation.

\subsection{Inter-observation Variability}
\fgl\ was found to be faint in radio, with the strongest detection being $\sim8\sigma$ in significance, and two epochs only providing upper limits on the radio flux density (Table ~\ref{table:obssum}). Our VLA observations detect \fgl\ at $23.6 \pm 4.8$\,$\mu$Jy and $47.7 \pm 6.0$\,$\mu$Jy at Epochs 1 and 2, respectively.  At Epochs 3 and 4, the source is undetected, with $3\sigma$ upper limits of 13.8\,$\mu$Jy and 16.2\,$\mu$Jy, respectively.  This demonstrates a clear variability in radio brightness of at least a factor of a few on timescales of days (see Fig 2 for radio interferometric images for each of the 4 epochs).  As mentioned above, strictly simultaneous radio/X-ray observations are needed to probe variability on the much shorter timescales on which \fgl\ is known to switch between X-ray low- and high-modes.

Given very low average radio flux density of \fgl, fitting for the spectral index calculated by imaging two base bands separately of radio emission from this source, or performing polarization imaging would not offer useful constraints. Indeed, the spectral index we obtain from the brightest epoch is $\alpha=0.7\pm1.0$, which is consistent with a flat spectrum, but with a large uncertainty.
Likewise, testing for short-timescale variability is strongly limited by the source brightness and the fact that the low modes -- where, in analogy to \psr, we expect radio brightening -- typically last for only a few minutes to half an hour at a time \citep{BH:2015}. Nevertheless, we attempted to search for such variability by splitting the observation into ten $\sim$3.2 minutes subintegrations; lightcurves thus obtained are presented in Fig. 3. We opted for 3.2 min, as it is also the average scan time of the source. In all epochs, except epoch 1, we find that the source variability is consistent with the rms noise of the image ($\sim 15$\,$\mu$Jy/beam per sub-integration). In the first epoch, we find a single bright peak of $71.2$\,$\mu$Jy/beam in one sub-integration, which is $\sim 3$ times brighter than the mean flux of the source in that epoch. However, due to the high rms noise, the significance of this bright peak is only $\sim 2\sigma$. This hints at a potential similarity of short term radio light curve behaviour of \fgl\ and \psr. To better investigate the short-timescale ($< 1$\,hr) variability of \fgl\, we performed (Gusinskaia et al. in prep) a strictly simultaneous radio/X-ray observation campaign, where the X-ray low/high-mode separation can guide the investigation of the radio light curve.\smallskip

\subsection{Radio Position}
Based on the brightest radio detection (Epoch 2), we used CASA's {\tt imfit} to fit a 2-D Gaussian constrained to have the same dimensions as the synthesized beam, which in this case had dimensions of $2.13^{\prime\prime} \times 0.87^{\prime\prime}$ at a position angle of $143^{\circ}$ (East of North).  In this way, we obtain a best-fit radio position (J2000):\\

\indent RA: $15^{\rm h}44^{\rm m}39.^{\rm s}3916 \pm 0.^{\rm s}0082$, \\
\indent Dec: $-11^{\circ}28^{\prime}04.^{\prime \prime}8497 \pm 0.^{\prime \prime}1556$.\\ 

Errors as they are currently reported above are statistical from fitting the image in plane.
This position is consistent to within 50\,mas of the optical counterpart (i.e. the neutron star's binary companion) GAIA DR2 position (J2000) of RA: $15^{\rm h}44^{\rm m}39.^{\rm s}388192$, Dec: $-11^{\circ}28^{\prime}04.^{\prime \prime}8735$ \citep{GAIADR2:Astrometric}, and thus establishes a convincing association. 

%USNO B1.0 catalog position (J2000) of  RA: $15^{\rm h}44^{\rm m}39.^{\rm s}38$ and Dec: $-11^{\circ}28^{\prime}04.^{\prime \prime}3$ cited in \cite{BH:2015}

\subsection{Radio-X-ray Variability}
To convert the radio luminosity to a frequency of 5\,GHz, for comparison with other LMXB systems, we assume a flat spectrum ($\alpha = 0$). This is consistent with our measurement as well as with both theoretical expectations for a partially self-absorbed synchrotron jet, and is what is observationally seen in most LMXBs. However, note that recent work by \citet{EF:2018} shows that populations of `radio-loud' and `radio-quiet' black hole X-ray binaries differ markedly in their spectral indices. This may point to a difference in the radio emission mechanism at lower radio luminosities and require more caution in assuming flat spectral index. While \psr\ showed short-timescale spectral variations during its minutes-long radio flares \citep{BDM:2018}, averaged over longer timescales (like those probed here for \fgl) the radio spectrum was still on average flat \citep[$-0.3 < \alpha < 0.3$;][]{DMM:2015}.  Thus, this extrapolation should be roughly correct. We find radio luminosities of $4.08\times10^{27}, 8.19\times10^{27}, 2.37\times10^{27}, 2.76\times10^{27}$\,erg~s$^{-1}$, at Epochs 1 to 4, respectively.  Figure~\ref{figure:lrlx} is a $L_R$ vs. $L_X$ diagram showing the derived radio luminosities at each VLA observing epoch, in comparison to the average \textit{Swift} X-ray luminosity quoted above.

%\textcolor{blue}{Pulsar search result inclusion here is pending}

\section{Discussion \& Conclusions}
\label{sec:disc}

%To conclusively classify a neutron star binary as a tMSP, one needs to demonstrate that the system has been observed in both the rotation-powered RMSP and accretion-powered LMXB states.  Nonetheless, in the absence of an observed transition, we can still make a strong case that a particular candidate system is of a similar ilk to the tMSPs and may in the future undergo a state transition.  Fortunately, the tMSPs provide a rich observational phenomenology that helps build such a connection (see \S1), with the caveat that there are only three such systems currently known and perhaps a larger sample will display a larger diversity.

Many of the key observational similarities of \fgl\ compared to the well-established tMSP systems were previously presented in \citet{BH:2015} and \citet{Bog:2016}. \citet{BH:2015} highlight the phenomenological similarity of \fgl's \textit{XMM-Newton} lightcurve to that of \psr\ and \xss\ where, during the accretion-powered state, the X-ray flux switches between quasi-stable `low' and `high' modes that are separated in X-ray luminosity by roughly an order of magnitude. Using the \textit{XMM-Newton} Optical Monitor (OM) and  MDM  observatory fast photometric observations, they saw that the system showed rapid optical variability, switching within a range $\sim 0.5$\,mag in a manner reminiscent of the limit-cycles seen in \psr. Furthermore, X-ray spectroscopy presented in \citet{Bog:2016} showed that the $1-80$\,keV emission (from joint \textit{XMM-Newton} and \textit{NuStar} data) can be modeled by an absorbed power law model with a photon index of $\Gamma \sim 1.7$. As with \psr, no significant spectral change was observed between the low and high X-ray mode \citep{BAB:2015}. This incredible similarity to \psr's X-ray spectroscopic behavior was also echoed by the similarity to its broadband spectral shape spanning optical, UV and $\gamma$-rays \citep[see Fig. 4,][]{Bog:2016}.

In the work we present here, we detect 10-GHz radio emission from \fgl\ varying in flux density from $47.7 \pm 6.0$\,$\mu$Jy down to $\lesssim 15$\,$\mu$Jy ($3\sigma$ upper limit) at four epochs spanning 3 weeks.  During this time \fgl's X-ray flux is consistent with the interpretation that it was in the persistent low-accretion-rate state seen in, e.g., tMSP \psr\ \citep{BAB:2015}.  Mapping \fgl\ in the radio/X-ray luminosity phase space (Figure~\ref{figure:lrlx}) shows that it populates a similar region to the well-established tMSPs \psr\ and \xss\ \citep{DMM:2015,BDM:2018,HSC:2011}.  This shared phenomenology adds further support to the previous classification of \fgl\ as a strong tMSP candidate, and suggests that the tMSPs truly have quite common observational properties across the electromagnetic spectrum. 

\fgl's radio flux density varies by a factor of $\gtrsim 3$ between epochs.  This is consistent with the behavior of \psr, where the radio flux density varies by a factor of two in $\sim 2$\,mins and by an order of magnitude in $\sim 30$\,mins \cite[see,][]{DMM:2015}.  However, since \fgl\ is at $3.8 \pm 0.7$\,kpc \citep{BSC:2017} instead of $1.368^{+0.042}_{-0.039}$\,kpc like \psr\ \citep{DAB:2012} it is much more challenging to study its radio variability on such short timescales.  Consequently, we plan to better investigate short-timescale ($< 1$\,hr) variability in a strictly simultaneous radio/X-ray observation campaign, where the X-ray low/high-mode separation can guide the investigation of the radio light curve.  These observations can determine whether, like \psr, \fgl's radio brightness increases during the X-ray low modes \citep{BDM:2018}. Given the general faintness of \fgl, this may require stacking many such mode instances over several hours.

%\subsection{\textcolor{red}{TMSPs and canonical LMXBs}}
In general, our findings support the idea that tMSPs as a class are surprisingly `radio-loud' at low X-ray luminosities, when compared to expectations based on previous radio/X-ray observations of NS-LMXBs at higher accretion rates \citep[e.g.][]{MF:2006,MJD:2011}. Even comparing \psr\ with other tMSP radio/X-ray detections, including the detection of \igr\ in outburst \citep[an LMXB-like outburst as seen in,][]{PFB:2013}, \citet{DMM:2015} tentatively postulated an empirical relation, $L_R \propto {L_X}^{0.7}$, between radio and X-ray luminosity for tMSPs. This speculative scaling relation is, perhaps coincidentally, very similar to that suggested for some black hole (BH) LMXBs \citep{GFP:2003, CNF:2003, Shaw:2021}, which show similar flat to inverted radio spectra arising from compact, self-absorbed jets \citep{BK:1979} over many orders of magnitude \citep{CCB:2013}. In the case of BH-LMXBs, the radiatively inefficient nature of the accretion flow can be attributed to the energy being liberated by in-falling matter either a) carried away in the outflow during the jet-dominated regime or b) not radiated but advected across the BH event horizon. However, while there may be some physical analogies to draw with BH jets, note that the neutron star magnetosphere and rapid rotation mean that the physical situation is likely quite different compared to tMSPs.

Placing \psr\ in the context of other neutron star (NS) LMXBs, previously \citet{MF:2006} established two proposed $L_R-L_X$ relations: 1) In Z-sources the radio luminosity was found to be proportional to ${L_X}^{0.66}$, whereas 2) In hard-state Atoll-sources and rapid burster NS-LMXBs the radio luminosity scaled as $L_R \propto {L_X}^{1.4}$. The relation found for Atoll-type sources remains valid over an order of magnitude in X-ray luminosity although jets being launched are less powerful than BHs at any given X-ray luminosity. They also found that after NS-LMXBs transition to the soft state their radio emission mechanism is not quenched, whereas in a recent study \citet{GDH:2017} find evidence for jet quenching in a NS-LMXB.  Furthermore, \citet{MF:2006} highlighted that, based on the observed steeper power law for NS-LMXBs, that they would never enter a jet-dominated state as some BH-LMXBs do. Also, \citet{MFK:2005} theorized that millisecond pulsars in X-ray binaries, given their relatively high magnetic fields \citep{Chakra:2005}, should produce jets with reduced radio luminosity.

More recent works, benefitting from a larger number of detected sources, show that the correlation for hard-state Atoll sources (index 1.4) should be viewed with skepticism given that this correlation was found over $1-1.5$ dex in $L_x$ \citep{TMP:2017, GDE:2018}. Indeed, the few sources tracked over more than $\sim 1-2$ dex in $L_x$ (mainly SAX~J1808.4$-$3658) show significant scatter in the radio/X-ray measurements, compared to the tighter correlation observed in BH-LMXBs such as GX$339$-$4$, V$404$ Cyg, or XTE J$1118+480$.  Furthermore, \citet{TMP:2017} question the very presence of single correlation. Such studies remain limited by the low number of detected NS-LMXBs, which are typically radio-fainter than BH-LMXBs by a factor of $\sim 20$ at a given $L_x$.

%\subsection{\textcolor{red}{TMSPs and AMXPs}}
While it may be tempting to connect the radio emission seen from tMSPs in the low-level accretion state (\psr, \xss, and \fgl) to that seen in the bright outburst of \igr\ (see thin dot-dashed blue line in Figure~\ref{figure:lrlx}) we caution that these plausibly represent different accretion regimes that are not connected by a well-defined radio/X-ray scaling. Indeed, based on the anti-correlated radio/X-ray moding they observed, \citep{BDM:2018} conclude that \psr\ does not fit a standard inflow-outflow model.  Further observations of tMSPs in the intermediate X-ray luminosity range of $10^{34} - 10^{36}$\,erg\,s$^{-1}$ are thus needed to better understand the similarities and differences at low and high accretion rates.  

%\textcolor{red}{JWTH: I'm going to check if there's something more we can say here.  No action from you required.} 

The class of accreting millisecond X-ray pulsars (AMXPs) provide an additional sample of sources that may help clarify this in the future.  Just as all three known tMSPs appear like the class of `redback' MSPs when they are in the rotation-powered state, some of the AMXPs appear like tMSPs in the accretion-powered state \citep[though none of these sources have shown radio pulsations during quiescence, e.g.][]{PJK:2017}.  \cite{TMP:2017} tracked various pulsating (IGR~J17511$-$3057, SAX~J1808.4$-$3658, and IGR~J00291+5934) and non-pulsating (Cen X-4) NS-LMXBs in outburst and down to quiescence. They showed that, in contrast to BH-LMXBs (which all seem to populate a radiatively inefficient track of $L_R\propto{L_X}^{0.7}$), NS-LMXBs exhibit more diverse radio/X-ray behaviors, possibly due to the neutron star's spin rate and magnetic field strength. For example, IGR~J00291+5934 (observed at $L_X = 10^{34} - 10^{36}$\,erg\,s$^{-1}$) displayed a tight disk-jet coupling behaviour with coupling index 0.7 in the $L_R-L_X$ relationship. Conversely, 
SAX~J1808.4$-$3658 showed highly-variable radio emission with strong radio emission epochs consistent with X-ray reflare tails.

Moreover, the recently discovered AMXP system IGR~J17591$-$2342 \citep{FBS:2018, SFR:2018,GRH:2020} was shown to be as radio loud as BH-LMXBs \citep[if at $d > 3$\,kpc]{RDW:2018}. Thus radio emission in some AMXPs is observed to be as bright as tMSPs. However, \citet{GDE:2018} show that pulsating (AMXP) and non-pulsating NS-LMXBs can only be fitted with a common slope in the $L_r$vs.$L_x$ diagram if tMSPs are excluded from the sample.Moreover, recent exploration \citep{EDR:2021} over an ensemble of AMXPs shows that neutron stars with weaker magnetic fields are usually radio fainter compared to stellar mass black holes. These works suggest a different physical situation in the case of tMSPs accreting at a low X-ray luminosity.  

Radio observations may also help in linking other X-ray binary systems to the tMSP class.  For example, \cite{DWR:2014} identified a peculiar X-ray binary, XMM~J174457$-$2850.3, close to the Galactic Center. This system periodically enters accretion-dominated outbursts ($10^{36}$\,erg\,s$^{-1}$) but at other times persists in an intermediate state where the X-ray flux is limited between $10^{33-34}$\,erg\,s$^{-1}$ for months at a time in a manner similar to tMSPs in long-term ($\sim$yr) low-level X-ray states.  Moreover, XMM J174457$-$2850.3's X-ray spectrum is well-fit by a power law with a photon index of 1.7. These observed properties are analogous to tMSPs and \citet{DWR:2014} invoked the neutron star's magnetic field interaction with the incoming accretion flow as a plausible reason for the long-term, low-level X-ray emission in such systems.

In the future, \fgl\ may eventually be detected to transition to a rotation-powered RMSP state or to enter a bright outburst at higher accretion rate, like that seen for \igr\ \citep{PFC:2013}.  Either situation could lead to the detection of the neutron star's rotation rate, via either radio or X-ray pulsations.  Such a detection would establish the rotational energy budget and magnetic field strength. %At present, Lovell radio telescope observations show no detectable radio pulsations. \textcolor{blue}{Analysis to be inserted.}  While \psr\ also shows no detectable radio pulsations during its accretion-powered state -- even when one searches separately during the low and high X-ray mode times \citep{BAB:2015} -- it appears that the pulsar wind is still operating during the low-level accretion state and that radio pulsations might in principle be generated but obscured by intra-binary material.  In this sense, the finding that \fgl\ may be a face-on system \citep{BSC:2017} could make radio pulsations more readily detectable in this case.  Thus, we feel that continued, deeper radio pulsation searches are warranted.

If \fgl\ transitions from LMXB to RMSP state, it will be only the fourth well established tMSP.  Currently, \fgl\ is the only (candidate) tMSP, apart from \psr, in the LMXB state. This makes \fgl\ a critical avenue in wihch to explore the nature of the tMSP accretion state and to compare it with the observations made for other transitional tMSPs in their current or previous LMXB state.

Lastly, our \fgl\ results strengthen the case that radio/X-ray/gamma-ray observations may provide a useful way of identifying new tMSP systems. Similar searches of select \textit{Fermi} un-associated objects and their multi-wavelength follow-ups have now yielded two other candidate tMSPs, \cxou and \fglnew, in sub-luminous disk state. Hence, we are targeting un-associated \textit{Fermi} gamma-ray sources using the VLA, Australia Telescope Compact Array (ATCA) and \textit{Swift} to look for sources whose position in the $L_R - L_X$ diagram is similar to the tMSPs.

\section*{acknowledgements}

We thank Tom Russell for discussions about the radio analysis. AJ acknowledges support from the \textit{NuSTAR} mission. AJ and JWTH also acknowledge funding from the European Research Council under the European Union's Seventh Framework Programme (FP7/2007-2013) / ERC grant agreement nr. 337062 (DRAGNET). AP acknowledges support from an NWO Vidi Fellowship. JCAM-J is the recipient of an Australian Research Council Future Fellowship (FT 140101082). SB was supported in part by NASA \textit{Swift} Guest Investigator Cycle 12 program grant NNX16AN79G awarded through Columbia University. 

We thank B.~Clark and the VLA schedulers for granting these DDT observations on short notice.  The National Radio Astronomy Observatory is a facility of the National Science Foundation operated under cooperative agreement by Associated Universities, Inc.  This research has made use of data and software provided by the High Energy Astrophysics Science Archive Research Center (HEASARC), which is a service of the Astrophysics Science Division at NASA/GSFC and the High Energy Astrophysics Division of the Smithsonian Astrophysical Observatory.

The data post-processing and results preparation for this work heavily relied on \textit{Common Astronomy Software Applications} (CASA 5.1) \citep{MWS:2007}. We would like to thank the staff at NRAO and externals involved in maintaining an extensive updated documentation and tutorials for CASA. This research also made use of Astropy, a community-developed core Python package for Astronomy \citep{Astropy:2013,Astropy:2018}). Other important softwares used in this research are Matplotlib \citep{Hunter:2007}, Seaborn and Scipy \citep{Scipy}. Lastly, we have made extensive use of the NASA Astrophysics Data System \citep{ADS:2000} and the arXiv e-print service \citep{Gin:2011}.

\bibliography{ApJ_ADJ}
\bibliographystyle{aasjournal_n}



\end{document}